# Assumptions Underlying Agile Software Development Processes

## Abstract

*Agile processes focus on facilitating early and fast production of working code, and are based on software development process models that support iterative, incremental development of software. Although agile methods have existed for a number of years now, answers to questions concerning the suitability of agile processes to particular software development environments are still often based on anecdotal accounts of experiences. An appreciation of the (often unstated) assumptions underlying agile processes can lead to a better understanding of the applicability of agile processes to particular situations. Agile processes are less likely to be applicable in situations in which core assumptions do not hold. This paper examines the principles and advocated practices of agile processes to identify underlying assumptions. The paper also identifies limitations that may arise from these assumptions and outlines how the limitations can be addresses by incorporating other software development techniques and practices into agile development environments.*

## 1. Introduction

As more organizations seek to gain competitive advantage through timely deployment of services and products that meet and exceed customer needs and expectations, developers are under increasing pressure to develop new or enhanced implementations quickly [15]. Agile software development processes were developed primarily to support timely and economical development of high-quality software that meets customer needs at the time of delivery. It is claimed by agile process advocates that this can be accomplished by using development processes that continuously adapt and adjust to (1) collective experience and skills of the developers, including experience and skills gained thus far in the development project, (2) changes in software requirements and (3) changes in the development and targeted operating environments. Examples of published agile processes are Extreme Programming (XP) [3][7][8][22][29][39], the Crystal process family, [13], SCRUM [33][34], Adaptive Software

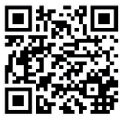

Development [19], and AUP (Agile Unified Process) [23] which has grown out of work on the UML [21][31][38].

Proper use of agile processes requires an understanding of the situations in which agile processes are and are not applicable. One way of determining whether an agile process is applicable in a particular situation is to check whether the assumptions underlying the process hold in that situation. If the assumptions do not hold then use of the agile process may not be appropriate. Prevailing descriptions of agile processes seldom present the underlying assumptions explicitly and thus it is difficult for developers and project planners to determine the applicability of agile processes to their projects and environments.

This paper identifies some of the assumptions underlying agile processes that can be used to help determine the applicability of agile processes to particular situations. The paper also discusses some of the limitations that may be inherent in agile approaches because of these assumptions. The assumptions were identified by examining published work on Extreme Programming (XP) [3][5], Scrum [34], the Agile Unified Process (as described by Craig Larman) [23], critiques of agile processes [10][27], and the principles stated by the Agile Alliance.

It is important to note that our critique of agile processes is concerned with identifying assumptions underlying a family of agile processes. Other critiques of agile processes have been published (e.g., see [10],[27]), but none of the critiques we have examined have focused on identifying assumptions underlying agile processes for the purpose of determining the scope of their applicability. For example, in the book "Questioning Extreme Programming" McBreen [27] presents a critique of XP in which he poses some important open questions and provides answers to other questions based on his personal experience, but he does not explicitly identify assumptions underlying agile processes. McBreen's critique was used as a source in our work along with other experience reported elsewhere (e.g., see [14], [24]).

The remainder of the paper is structured as follows. In section 2 we give an overview of a typical agile process, eXtreme Programming, to give the reader a concrete example of an agile process. In section 3 we describe the assumptions that we have identified. In section 4 we identify some of the limitations that arise in situations in which these assumptions are not met and suggest how they can be addressed by adapting some of the agile process techniques and practices. We conclude in section 5 with an overview of the results of our work and an outline of issues that require further investigation.

## 2. Overview of eXtreme Programming – A Representative Agile Process

There are a variety of software development processes that currently claim to be agile. Space does not allow us to give an overview of all of the agile processes we have reviewed. However, since Extreme Programming (XP) is probably the most well-known agile process, we use it to illustrate representative agile process concepts.

## *Extreme Programming (XP)*

It can be argued that the popularity of XP helped pave the way for other agile processes. Kent Beck, one of the chief architects of XP, states that XP is a "lightweight" development method that is tolerant of changes in requirements. It is "extreme" in that "XP takes commonsense principles and practices to extreme levels" [5, p. xv].

XP is based on the following values:

- *Communication and Feedback*: Face-to-face and frequent communication among developers and between developers and customers is important to the "health" of the project and the products under development. Feedback, through delivery of working code increments at frequent intervals, is also considered critical to the production of software that satisfies customer needs.
- *Simplicity*: XP assumes that it is more efficient to develop software for current needs rather than attempt to design flexible and reusable solutions. Under such an assumption, developers pursue the simplest solutions that satisfy current needs.
- *Responsibility:* The responsibility of producing high-quality code rests ultimately with the developers.

XP consists of technical and managerial practices that are integrated in a complementary manner. The architects of XP take great care to point out that the individual techniques and practices of XP are not new; it is the manner in which they are woven together that is unique. They also stress that the techniques and practices have proven their worth in industrial software development environments.

### XP Process and Practices

The four core activities of XP are (1) coding, (2) testing, (3) listening to the customer and to other developers, and (4) designing as an implicit part of the coding process. XP encourages an informal design specification process in which developers discuss solutions by sketching informal models on some presentation medium (e.g., whiteboard, flip chart). These models are created primarily to help developers understand and communicate ideas during development, and are not intended to be precise descriptions of the solution.

In order to support the five fundamental principles of XP – namely *rapid feedback*, *simplicity*, *incremental changes*, *embracing change*, and *quality work* – XP offers a number of practices. The early accounts of XP [7] offered twelve practices, but since then the number of practices has increased (see [26],[40]). We give an overview of some of the original practices in what follows.

*Pair programming,* one of the more well-know XP practices, is a technique in which two programmers work together to develop a single piece of code. The two programmers typically work together at one computer, collaborating to design, implement and test a software solution (program) [18][41][42]. At any point in time one programmer is directly working on the code, while the other observes, provides alternative approaches, acts as a reviewer and provides instant feedback. The two programmers switch their roles often, sometimes even after just a few minutes. This approach has been shown to yield significantly higher productivity and code quality than is achieved by two programmers working separately [40]. The intent is that two programmers working and evaluating the code and design are likely to complement each other's skills, continually propose and evaluate alternatives and are more likely to recognize errors in the code while it is being developed [41, p. 328]. Pair programming is based on two assumptions: (1) active reviews are the most effective way to detect errors, and (2) different people see a problem from different perspectives and will thus have a combined approach to problem-solving that is more effective than individually applied approaches.

*Refactoring*, *unit and acceptance tests*, *collective code ownership*, and *continuous integration* together tackle the problem of evolving code during XP-based development. Refactoring occurs when a change to the internal structure of a system preserves the externally observable functionality of the system. Refactoring is especially effective when large changes can be decomposed into smaller steps that can be carried out using refactorings that have been developed by Fowler and others [17]. These refactorings can be viewed as code transformation patterns, and their use allows one to reduce the task of validating code after a complex change to validation of smaller change steps.

After a refactoring, tests are run to ensure that parts that should not be affected by the changes are intact and that the changes are implemented correctly. Collective code ownership allows developers to appropriately change parts of the code that they did not write in order to implement a change, while continuous integration allows developers to demonstrate the current status of development more frequently.

## XP: An Assessment

Although XP is considered an extreme process it is not devoid of rigor. In particular, XP's focus on code should not be interpreted as an endorsement of code "hacking". XP stipulates that developers follow all its practices in order to realize the benefits of agile development. As has been pointed out by McBreen and others [27], it takes enormous discipline to apply XP and, for this reason, some projects may find it difficult to adopt an XP-compliant process.

A significant problem with XP is its reliance on source code for documentation. This usually leads to situations in which in-depth knowledge of software products (e.g., design rationale, trade-off considerations) exist only in the heads of the developers who developed the products.

Loss of these developers could lead to significant organizational memory loss that could impair an organization's ability to complete projects in a timely manner.

XP specifically targets small- to medium-sized projects. XP proponents claim that XP's unique composition of best practices, and its omission of time-intensive software engineering activities (e.g., detailed specification or modeling of requirements and design), can help downsize otherwise large projects. There have also been proposals for scaling the XP process to large projects (e.g., see [14]), including an approach that involves hierarchically structuring XP and installing a steering committee to guide the individual projects [20].

To date, there are few objective surveys of projects claiming to use XP. One such survey [32] was conducted on 45 projects that were labeled as XP projects by the developers. The results show that XP is still in the "hype phase": it was not clear whether the claimed successes were based on developer enthusiasm or on the XP practices. A summary of the survey results is given below:

- More than 90% of the projects claimed to be successful (as judged by the developers, not by the customers)
- All surveyed said they would like to use XP again. None blamed failures on XP.
- The unavailability of customers was frequently the highest risk identified.
- Use of unit tests and pair programming were considered important practices.
- 33% used XP because it seemed more attractive than alternatives; 28%, because it fit the project requirements best; and 9% because the management or customer wanted it.

The results of this survey also indicate that there may be situations in which the basic assumptions underlying XP are valid. XP assumes that the cost of change slowly approaches some limit over time, rather than increasing exponentially as has been traditionally assumed [7]. XP practices are based on the assumption that correcting requirements errors and design flaws later does not cost significantly more than if they were detected and removed earlier. This assumption allows developers to do less than thorough analysis and design in the early phases and, instead, make improvements throughout the course of the project by refactoring the code. There is no objective evidence that this assumption is valid in general, but it can be argued that the cost of change curve can be flattened by using reusable design experiences in the form of architectural and design patterns, and capitalizing on new technologies supporting rapid program development (e.g., libraries, components and frameworks, and more powerful compilers that enable short and incremental compilations).

In the testing area, an issue that XP practitioners face is determining the tests needed to adequately cover the code. It has been recognized by some advocates that knowledge of systematic testing techniques can be beneficial when developing unit and acceptance tests in XP [27].

The set of tests developed for an application can be viewed as a model of the system: it describes an exemplar set of data with intended behavior. The tests are not necessarily readable by customers, but developers can use the tests to gain understanding of code they did not write, by exercising the code using the tests. This implicit model of the system is a necessary prerequisite for collective code ownership and refactoring techniques. However, if the rationale behind a test is not documented, over time it may become unclear what aspects are being tested.

## 3. Identifying Assumptions Underlying Agile Processes

In recent years a number of processes claiming to be "agile" have been proposed in the literature. To avoid confusion over what it means for a process to be "agile", seventeen methodologists and proponents of agile processes met to discuss and come to an agreement on what "agility" means. The result of the meeting was the formation of the *Agile Alliance* and the publication of a manifesto that included a list of principles agile processes should support [1]. A summary of these principles (numbered and in order as reported by the Agile Alliance [1] are given in Figure 1 below.

| | |
|---|---|
| 1. | "Our highest priority is to satisfy the customer through early and continuous delivery of valuable software." |
| 2. | "Business people and developers must work together daily throughout the project." |
| 3. | "Welcome changing requirements, even late in development." |
| 4. | "Deliver working software frequently." |
| 5. | "Working software is the primary measure of progress." |
| 6. | "Build projects around motivated individuals. Give them the environment and support they need, and trust them to get the job done." |
| 7. | "The best architectures, requirements, and designs emerge from self-organizing teams." |
| 8. | "The most efficient and effective method of conveying information to and within a development team is face-to-face conversation." |
| 9. | "Agile processes promote sustainable development." |
| 10. | "Continuous attention to technical excellence and good design enhances agility." |
| 11. | "Simplicity is essential." |
| 12. | "Project teams evaluate their effectiveness at regular intervals and adjust their behavior accordingly." |

Figure 1: Principles of the Agile Alliance

The manifesto of the "Agile Alliance" is a condensed definition of the values and goals of "Agile Software Development" and is detailed through these principles which can be viewed as a set of policies and rules that should be supported by processes claiming to be "agile". These principles provide a good base for identifying assumptions underlying agile processes. In the next section we review these principles and identify assumptions that appear to be made when accepting these principles.

## Principles, Practices, Assumptions, and Limitations

Figure 2 summarizes our view of the relationship between Principles, Practices, Assumptions, and Limitations. We have taken the view that there are assumptions, usually unstated, that led to the acceptance of the Agile Alliance's principles. There are also assumptions made by developers (again unstated) regarding what these principles mean and their relative importance. Based on these principles and assumptions, development practices are set in place. Whether intended or not, these assumptions lead to limitations in the resulting agile processes. The limitations discussed later in this paper are based in part on our assessment of the assumptions that exist behind agile process principles and practices. In this section we identify assumptions underlying the Agile Alliance principles, as we perceive them. The discussion in this section is organized around clusters of related principles, where each cluster gives rise to a distinct set of assumptions. We also identify examples of situations in which the assumptions may not hold.

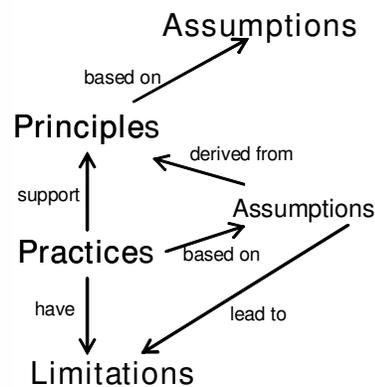

Figure 2: Relationships Between Principles, Practices, Assumptions, & Limitations

Principle 1. "Our highest priority is to satisfy the customer through early and continuous delivery of valuable software."
Principle 4. "Deliver working software frequently."
Principle 5. "Working software is the primary measure of progress."

The prominence of Principle 1 serves to remind developers that software is developed to perform services that add value to users at the time of delivery. (Principles 4 & 5 can be viewed as consequences of principle 1, and thus are discussed together here.) Developers and project planners need to keep in mind that customer needs evolve through use of systems. Support for Principle 4 allows developers to gauge and address evolving customer needs. Agile processes provide support for these principles by structuring development activities into short fixed-time iterations that each produces working code. These fixed-time iterations make agile processes predictable along the time dimension. The price paid is that product scope can be unpredictable: in meeting an iteration deadline, developers can choose not to implement features originally marked for implementation in the iteration. Agile processes utilize practices that help developers minimize the time it takes to realize elicited requirements in working code. In XP, these practices include building simple designs, continuous integration, collective code ownership, and refactoring.

The frequent delivery of working code gives the project visibility in terms that the customer can relate to (i.e., in terms of an evolving executable product), rather than in terms of evolving plans in the form of requirements and design documents that are often not presented in terms a customer can relate to or understand. Customers can use the increments delivered by iterations as the basis for (1) determining project progress as specified by Principle 5 and (2) clarifying and refining requirements. Short iteration lengths facilitate timely customer feedback that can help ensure that the end product will meet customer needs at time of delivery.

The importance of involving end-users and customers in the software development process is widely recognized, and was the primary motivation of work in the early 1980s on "end-user" software development [12]. Much of this early work focused on developing mechanisms that would allow end-users to directly contribute to the development of requirements and designs, and understand the artifacts created by software designers. The work at that time focused on developing requirements and design notations that are "customer-friendly", that is, that can be used to create requirements specifications and designs that provide customers with significant insight into requirements and designs. Mechanisms such as fourth-generation programming languages (4GLs) and executable requirements [4][30][35][36] and design specifications were considered to be enabling technologies for end-user software development. Rather than emphasize technical facilitators of customer interaction, agile processes emphasize continual collaboration of developer and customer teams.

## Assumptions Underlying Principles 1, 4, and 5

*Visibility Assumption*: Project visibility can be achieved solely through delivery of working code.

Visibility of software development projects is traditionally accomplished through reports, specifications, and measures of quality and productivity, and the working application (code) is only seen after the developer has done a large amount of work and spent a great amount of time working on the project. For customers, it is easier to gain a sense of whether the project is progressing in the direction needed if they can actually *see* the user interface and actually *see* the software *do* the things they need it to do, rather than simply relying on reports, specifications, and other measures. Because it is the end-product the customer really cares about, the primary measure of progress in agile processes is based on the code developed in the project.

This works well for software that is equipped with user interfaces that evolve over time. In projects in which the user-interface is not always part of a deliverable increment, or projects concerned with developing systems with no human interface (e.g., embedded systems) some other means of visibility is needed. For such projects, software simulations, coverage of acceptance tests, and formal reviews and inspections of deliverable increments can provide some visibility.

*Iteration Assumption*: A project can always be structured into short fixed-time iterations.

Agile processes require developers to group required features into loosely coupled bundles that can each be addressed in short, fixed-time iterations. Such decomposition is usually based on an implicitly imposed architecture consisting of loosely-coupled modules and is thus desirable. The assumption here is that the developing application can be broken into small, discrete increments, that can be developed and demonstrated in short fixed-time intervals, and that after each of these iterations the customer will be able to observe additional functionality in the product. Because of this, the customer will be able to give frequent feedback as to the progress of the project, indicating whether it is being developed as needed / expected or not.

Structuring work in small bundles that can be implemented quickly may not always be possible. For example, in some complex systems an application may be required to interface with a number of subsystems in complex ways just to provide basic services that are of value to customers. In these situations it may not be possible create small enough bundles of features to tackle in an iteration because of the tight dependencies.

## Principle 2. "Business people and developers must work together daily throughout the project."

The interaction between developers and end-users in agile processes is concerned primarily with resolving feature-related issues and determining the scope of effort. This interaction does not occur only at the start of a project; it occurs throughout the process. Specifically, agile processes advocate interactions that could involve customers (1) providing inputs in the form of informal descriptions of expected behavior (for example, stories in Extreme Programming), (2) answering questions about desired features, (3) collaborating with developers in resolving issues pertaining to features to be implemented, and (4) collaborating with developers to evolve project plans

One of the most important effects of this close collaboration between users and developers is the better understanding of each other's problems and needs, which reduces human interaction problems and thus significantly enhances the chance for a successful project result.

This principle is not only applicable to the interactions between developers and customers; it also extends to interactions among developers as well. Frequent interaction allows developers to quickly resolve problems and misunderstandings, and to more quickly and reliably move forward on the project.

## Assumptions Underlying Principle 2

*Customer Interaction Assumption*: Customer teams are available for frequent interaction when needed by developers.

Some major assumptions here are that the customer is available at the time the developers need to interact with them, and that the customer can always reschedule other work so that there is time for frequent interactions with the developers. The reality is that it may not always be possible for a customer to reschedule other work.

*Team Communication Assumption:* Developers are located in time and place such that they are able to have frequent, intensive communication with each other.

This assumption is very similar to the Customer Interaction Assumption, but is focused on the ability of developers to interact with each other. Just as the Customer Interaction Assumption assumes a certain amount of common time, place, resources, and availability, so does the Team Communication Assumption. Time, place, resources, and availability must all be coordinated and provided in order to allow this principle to be supported. Examples of projects in which this assumption does not hold are plentiful. It is not unusual to have development teams on a single project that are dispersed in wide geographical areas involving many time zones.

## Principle 8. "The most efficient and effective method of conveying information to and within a development team is face-to-face conversation."

In agile processes, face-to-face communication is emphasized over formal and precise documentation, but also over tele-/video-conferences or email conversations. The agile process community claims that more is gained through informal personal communications than through communication based on formal documentation, even though the ability to track all information disappears. An advantage of face-to-face communication is that the parties involved can change the direction of the discussion as needed to gain insights into the topic under discussion, and can observe and respond to non-verbal communication cues as developers and/or customers interact.

Even though formal and technical communication mechanisms are discouraged, protocols or "to do" lists should be used to keep track of things that have been discussed.

## Assumptions Underlying Principle 8

*Face-to-Face Assumption*: Face-to-face interaction is the most productive method of communicating with customers and among developers.

It is hard to imagine Principle 8 being realized without having co-located customers and developers and without schedules that allow frequent interaction during a project. If customers and developers are all co-located, even to the extent they can walk down the hall and talk with each other, then any time questions, issues, or problems arise, they can be addressed immediately and clear resolution may be immediately obtained. Without face-to-face contact there is increased potential for mis-communication, and there is always the difficulty of getting in contact – the "telephone tag" or "e-mail tag" problem.

The de-emphasis of documentation as a communication aid is based on an assumption that tacit knowledge is to be valued over externalized knowledge. Proponents point out that developers need to internalize externalized knowledge to make it useful and that learning can be accomplished by sharing of tacit knowledge through conversations [27]. Critics have argued that the focus on tacit knowledge makes projects that use agile processes dependent on experts [10]. Another concern is that valuing tacit knowledge over externalized knowledge can lead to corporate memory loss and a reduced ability for an organization to systemically learn from its collective experience. An organization that is concerned with its ability to effectively learn from past experience needs to value both tacit and externalized knowledge and understand their interactions. Tacit knowledge is critical to building externalized knowledge, as pointed out by Nonaka [25][37], and externalized knowledge can interact with tacit knowledge to reveal hidden or create new tacit knowledge. Organizations that value systemic learning need to foster environments that not only support the sharing of tacit knowledge but also support externalizing tacit knowledge.

*Documentation Assumption*: Developing extensive (relatively complete) and consistent documentation and software models is counter-productive.

Given agile developers' code-centric focus (see principles 1, 4, & 5 above), this downplaying of documentation and software models is not surprising. The assumption is that it is more reliable to determine specifications and design from code than from other documents – especially since specifications, requirements, design documents, and models may not be kept up-to-date when code is changed. Thus, the code is the most accurate and reliable description of what a system does and how it was designed.

A reason for the agile process community's disenchantment with modeling may be a result of prior experiences with commercial modeling tools that were nothing more than tailored drawing environments. Such tools provided very little support for the more difficult tasks of maintaining (1) traceability links across models and (2) consistency between models and their implementations. Current modeling tools have moved somewhat beyond this and now provide support for code generation and round-trip engineering. More importantly, major tool vendors are currently extending their offerings to support the Object Management Group's model-driven software development approach, known as the Model-Driven Architecture (MDA) [28]. MDA is based on a separation of platform-specific system details from platform-independent details. MDA-based tools provide mechanisms for mapping platform-independent details to platform-specific implementations, with a significant portion of the mapping being automated through the use of patterns, templates, and other forms of reusable experiences. In the MDA approach, models are the central artifacts, and the use of MDA tools can help speed up development through automated generation of significant portions of application and middleware code and by raising the level of abstraction at which developers work.

For customers who contract with developers to provide systems, precise models may not be necessary. However, there are situations in which models are valuable in their own right, and in which it would be beneficial to maintain these models for future use. Some of these situations are described below:

- Evolving large complex systems that have long life cycles: The availability of good models can reduce the cost and effort of modifying such systems. Without these models, developers are forced to analyze source code to understand it and determine the impact of change. Studies have shown that a significant portion of the effort required to evolve systems is spent understanding the code. Good models can help ease this task.

- Managing enterprise systems to ensure alignment with business goals: Good models of business processes and systems can be used by enterprise architects to (1) check that planned and implemented systems align with business goals, (2) identify how existing systems services can be composed to create new services, (3) identify redundancies in systems (particularly when organizations or sub-organizations merge with other organizations or sub-organizations), (4) identify reusable development experiences, and (5) determine the impact of change on existing systems. Business- and system-level models, well-defined mappings between them, and the correspondence with code, can greatly enhance the management of enterprise systems.

Good models and documentation can also be used to bring new hires up to speed on the business and the systems being developed, and help component users determine whether a software component really addresses their requirements.

## Principle 3. "Welcome changing requirements, even late in development."

Requirements will change during software development to reflect changes in (1) the environment in which the software will be implemented, and in (2) the development environment. This has been widely recognized (within and outside of the agile process community) and is one major reason for rejecting the simple waterfall model. Evolving requirements is often viewed as an inherent problem of software development. The agile process community views requirements changes as providing opportunities for evolving software that can enhance the customer's competitiveness in a rapidly evolving environment. Development teams that can handle such changes and produce software that is useful to the customer at the time of delivery (rather than at the start of the project) are more likely to have satisfied customers. Short iteration cycles and the "plan one iteration at a time" approach are claimed to provide the flexibility needed to accommodate changes in agile processes. Agile process proponents claim that adhering to this principle significantly increases the competitiveness of a company.

## Assumptions Underlying Principle 3:

*Changing Requirement Assumption:* Requirements always evolve, because of changes of technology, customer needs, business domains or even acquisition of new customers.

The assumption here is basically a re-statement of the principle. Changing requirements are not regarded as necessarily bad, but are welcomed as an opportunity to satisfy customer needs even better than when inflexibly sticking to old requirements. If customer needs change late in the project, then making sure that the project adapts to these changes is important to making the project a success.

*Cost of Change Assumption*: Cost of change does not dramatically increase over time.

Agile processes challenge the widely-accepted belief that errors introduced early and detected late in the process have significantly higher costs than errors detected early. Agile process proponents argue that appropriate use of new development technologies and practices can reduce the cost of uncovering errors late in the development process. One can make a credible case that the use of technologies and practices such as (1) very fast compilers with sophisticated type systems, (2) integrated development environments, (3) systematic improvement of code through refactoring, and (4) automated test suites can help manage the cost of detecting and removing errors even when the errors are uncovered late in the process. It is also clear that the cost of correcting errors that can be fixed by localized changes – that is, changes with limited impact – should be relatively stable over time. On the other hand, it is also clear that certain types of errors – for example, architectural design flaws that seriously compromise the integrity of the design, or errors that require corrective actions that have wide impact – are more costly to correct the later they are uncovered.

Principle 6. "Build projects around motivated individuals. Give them the environment and support they need, and trust them to get the job done."
Principle 7. "The best architectures, requirements, and designs emerge from self-organizing teams."
Principle 12. "Project teams evaluate their effectiveness at regular intervals and adjust their behavior accordingly."

Agile processes such as XP and Scrum emphasize the need to shelter developers from distractions so that they can focus solely on project activities. Management's role is to facilitate development by ensuring that developers have the resources they need when needed, and that they are not distracted by concerns outside the scope of the project. Management should also refrain from imposing and micro-managing the development team: developers should be trusted to get the work done using a process that is based on their collective experiences (i.e., the team should be self-organizing). Motivation is one of the most important properties humans need in order to achieve ambitious goals with good quality results.

It can be difficult to transform a traditional team into an agile, self-organizing team. In some agile processes this can require team leaders to transfer some of their traditional responsibilities to team members. The short iterations of agile processes allow the project leader to test transfer of responsibility, and thus incrementally build trust in a team's ability to get the job done.

It is claimed that support for Principle 6 leads to products that are of higher quality, meet customer requirements at delivery time, are better structured, and require less effort to build than those created using more predictive (heavy-weight) processes. However, we are not aware of any empirical studies that provide evidence of improved quality and reduced effort as a result of using agile processes.

The frequent reviews advocated by agile processes focus on the products and the process used to develop the products. The planning of iterations also allows for reflection on previous results and adjustment of future iterations. As the customer is continuously involved, different viewpoints on the effectiveness of the project team can be obtained and flexible reaction to this reflection is possible. The agility in agile processes is achieved through self-examination of the processes used and corresponding adaptation of the process.

A self-evaluation and adjustment of a project, however, needs a project environment that allows flexible adaptations. If the environment is "hostile", this means it is inflexible to change, its customers are not willing to actively participate, its contractors insist on written specifications to be fulfilled, etc. It becomes much more difficult to act in an agile manner.

## Assumptions Underlying Principles 6, 7, and 12

*Team Experience Assumption*: Developers have the experience needed to define and adapt their processes appropriately.

Another way of saying this is that an organization can always form a team consisting of bright, experienced problem solvers capable of evolving their process effectively. A development team that (1) consists of developers with solid programming skills and relevant process and product experience, and (2) has the ability to converge through rational discussions will likely be able to effectively define and adapt their project processes. Unfortunately, not all development teams have these qualities. Some need guidance in determining appropriate processes. For such teams, a "standard" process may work better than an adaptable process that they could find difficult to control. Indeed, the Team Experience Assumption is critical to the success of agile development projects.

It is generally accepted that there is no single process that will be applicable to all projects. On the other hand, there are a number of best practices, techniques, and experiences that developers can use in appropriate situations. Software development teams that consist of leading members that understand the situations in which particular processes and practices are applicable are more likely to be successful within an agile environment. It is therefore the responsibility of future agile developers to develop such an understanding by gaining experiences with a variety of approaches. Teams consisting of developers with these skills are more likely to benefit from the use of agile processes.

Embedded within the Team Experience Assumption there seem to be two more assumptions: The Self-Evaluation Assumption and the Self-Organizing Assumption.

*Self-Evaluation Assumption*: Teams are able and willing to evaluate themselves.

A team must evaluate its process if it hopes to be able to adapt and/or improve the process. The assumption is that the team is able and willing to do this. This is difficult in a project culture, where less than optimal behavior is regarded as a serious liability, and thus team members may be reticent to rive honest self-evaluations. Furthermore, even if the team is willing to self-assess, the team also needs to have the skills to do so. This basically boils down to the necessity of the team members having gained experience in previous successful projects to be able to compare this project's effectiveness with previous ones and identify possible improvements.

*Self-Organization Assumption:* The best architectures, requirements, and designs emerge from self-organizing teams.

The assumption here is that not only are the best architectures, requirements-elicitation, and designs produced from self-organizing teams, but that the resources exist for self-organizing teams to be created, and that management allows and supports this approach.

While this assumption is basically a restatement of Principle 7 it should not be regarded as simply redundant. It is assumed that teams will self-organize, drawing from the most highly-qualified talent-pool available, thus creating teams of diverse capabilities, and thus the ability to create the best products possible. The concept of self-organizing teams is very different from how many organizations work. Thus, if an organization expects to gain the most from applying Agile processes it should be aware that its management of teams may need to be radically re-designed.

## Principle 9. "Agile processes promote sustainable development."
## Principle 10. "Continuous attention to technical excellence and good design enhances agility."

Using agile processes, developers focus on delivering just the functionality needed and timely evolution of the software in response to changes in customer needs and the market. Agile process advocates stress the importance of fostering a development environment that continually stimulates and motivates developers. Rules, such as XP's 40 hour weeks and No-Overtime, target this principle.

The primary quality control activities in agile processes are code testing and customer feedback. Frequent review meetings are advocated in processes such as Scrum, while Extreme Programming advocates continuous reviews through pair-development of code. Extreme Programming also advocates the building of test cases before the building of code, and the use of regression tests to ensure that implemented changes do not have undesirable effects.

As systems grow through time, an initially well-designed architecture may become increasingly blurred. Extreme programming uses the refactoring technique to constantly redesign the system and therefore keep the design quality at an optimum. This keeps implementations enhanceable for further iterations and maintainable for the future.

## Assumptions Underlying Principles 9 and 10

*Quality Assurance Assumption*: Evaluation of software artifacts (products and processes) can be restricted to frequent informal interviews, reviews and code testing.

XP replaces the traditional review with pair programming, collective code ownership, and a rigorous "test first" approach. These approaches provide opportunities for continuous review and improvement of the product during development. Scrum and Crystal advocate the frequent use of workshops, review meetings, and interviews to evaluate products and the process, and use the results to adapt the process accordingly.

Despite their apparent strengths, it seems that the informal evaluation techniques of agile processes may not be sufficient for establishing the quality of safety-critical systems – systems in which in which failure can result in direct injury to humans or cause severe economic damage. Development and testing techniques which are more formal and/or rigorously planned may help ensure the quality of these types of systems. Those, however, require significantly more effort and are thus a lot more expensive. Validating an implementation against its requirements through analysis techniques, for example, means that a precise and detailed specification model must be derived from the requirements.

*Continuous-Redesign Assumption*: Systems can be continuously redesigned (refactored) and still maintain their structural and conceptual integrity.

One major assumption behind agile development is that a design can and should be continuously redesigned. Day after day the design is re-evaluated, and as better designs are determined, refactoring and re-development are carried out. Of course, a big assumption for this is that this redesign can be carried out for a significant amount of time without destroying the structural and conceptual integrity of the design and the product.

## Principle 11. "Simplicity is essential."

This principle is a direct reaction to what is perceived as unnecessary complexity imposed by heavyweight processes. Agile processes therefore advocate simplicity both in the code and the tools used. Code generators or frameworks are advocated only if they provide clear value to the project. Of utmost importance, the design is to be kept simple to support future iterations. Therefore, a focused architecture satisfying today's needs is preferred to a general architecture that is "designed for the future". This follows the idea that future changes are almost absolutely unforeseeable and it therefore makes little sense to plan for a future that might not happen. Furthermore, redesign is encouraged if it simplifies the system and removes unneeded functionality.

*Application-Specific Development Assumption*: Reusability and generality should not be goals of application-specific software development.

Part of keeping an application simple is to stay focused on current requirements and needs rather than trying to build a more general system that will "more easily be adapted to future needs". Building a more general and "adaptable" system tends to make the system more complex.

Agile processes encourage the use of reusable artifacts (e.g. design frameworks, patterns) only when it is clear that their use can help reduce costs or increase quality. Building a generalized piece of code (one that can be used in a number of situations) is encouraged in agile processes when it is clear that such generality can be used in the same project (e.g. factoring common method parts). Many agile process advocates claim that a focus on creating general solutions can result in efforts on making systems amenable to changes that may never occur. This is true especially of those developers who adopt the XP approach to agile development; it is not necessarily inherent in the principle itself. By focusing on building software that implements the specific requirements at hand, and keeping this well-designed, agility for completing this development is enhanced.

Part of this assumption is the idea that the long-term costs of development are smaller if at any given time the focus is on current requirements rather than on generalization. Of course, this assumption is debatable, since it may turn out that if the original design had been more general, it would have been easier, and thus less costly, to add and adapt features over time. But this viewpoint must be held in contrast to the view that it is hard to know what future changes will be required, and thus that developers may be investing in generalizations that will never be needed.

*Continuous-Redesign Assumption (re-iterated)*: Systems can be continuously redesigned (refactored) and still maintain their structural and conceptual integrity.

Generally, when a system is first designed it is in its simplest state. Over time, and after many changes have been made, the design typically degrades and thus the system becomes more "complex". The assumption here is that this continuous re-design actually keeps the system simpler.

Figure 3 below summarizes the assumptions identified in this section that lie behind the principles of the Agile Alliance.

| 1. The Visibility Assumption | Project visibility can be achieved solely through delivery of working code. |
|---|---|
| 2. The Iteration Assumption | A project can always be structured into short fixed-time iterations. |
| 3. The Customer Interaction Assumption | Customer teams are available for frequent interaction when needed by developers. |
| 4. The Team Communication Assumption | Developers are located in time and place such that they are able to have frequent, intensive communication with each other. |
| 5. The Face-to-Face Assumption | Face-to-face interaction is the most productive method of communicating with customers and among developers. |
| 6. The Documentation Assumption | Developing extensive (relatively complete) and consistent documentation and software models is counter-productive. |
| 7. The Changing Requirements Assumption | Requirements always evolve, because of changes of technology, customer needs, business domains or even acquisition of new customers. |
| 8. The Cost of Change Assumption | Cost of change does not dramatically increase over time. |
| 9. The Team Experience Assumption | Developers have the experience needed to define and adapt their processes appropriately. |
| 10. The Self-Evaluation Assumption | Teams are able and willing to evaluate themselves. |
| 11. The Self-Organization Assumption | The best architectures, requirements, and designs emerge from self-organizing teams. |
| 12. The Quality Assurance Assumption | Evaluation of software artifacts (products and processes) can be restricted to frequent informal interviews, reviews and code testing. |
| 13. The Application-Specific Development Assumption | Reusability and generality should not be goals of application-specific software development. |
| 14. The Continuous-Redesign Assumption | Systems can be continuously redesigned (refactored) and still maintain their structural and conceptual integrity. |

Figure 3: Summary of Assumptions Behind Principles of the Agile Alliance

## 4. Tackling Limitations of Agile Processes

From the discussion in the previous section it should be clear that the assumptions underlying agile processes do not hold in all software development projects and environments. This should not be surprising: Agile approaches are not process silver bullets. Because these assumptions are not met in all organizations and/or development environments, agile approaches, in their current forms, do have limitations. It is possible to extend agile processes to address their limitations. Such extensions can involve incorporating principles and practices often associated with more predictive, plan-based, or "traditional" development processes into agile processes. In general, users of agile processes need to ensure that practices based on assumptions that are not valid in their development environments are modified accordingly.

In this section we identify some limitations associated with the assumptions made by agile processes and discuss how some of these limitations can be addressed. For each limitation we characterize the situations in which the assumptions that lead to the limitation do not hold and discuss how agile processes can be modified to extend the applicability of agile processes. Not all the assumptions identified in the previous section lead directly to limitations discussed in this section.

Figure 4 below summarizes the relationships between the limitations discussed in this section and the relevant assumptions identified in the previous section. We have identified two categories of limitations: Personnel-related limitations and Product-related limitations. The assumptions that are people-oriented tend to lead to limitations in the Personnel category, while assumptions about the types of artifacts produced in a project lead to limitations in the product category.

| Assumptions | Agile Process Limitations | | | | | |
|---|---|---|---|---|---|---|
| | Personnel Limitations | | | Product Limitations | | |
| | Limited support for distributed development environments | Limited support for subcontracting | Limited support for development involving large teams | Limited support for building reusable artifacts | Limited support for developing safety-critical software | Limited support for developing large, complex software |
| Customer Interaction Assumption | X | X | X | | | |
| Team Communication Assumption | X | X | X | | | |
| Face-to-Face Assumption | X | X | X | | | |
| Changing Requirements Assumption | | X | | | | |
| Documentation Assumption | X | X | X | X | X | X |
| Quality Assurance Assumption | | | | X | X | X |
| Iteration Assumption | | | | | | X |
| Application-Specific Development Assumption | | | | X | | |
| Continuous Redesign Assumption | | | | X | X | X |

Figure 4: Limitations of Agile Processes and Related Assumptions

## *Limited Support for Distributed Development Environments*

Distributed development environments are environments in which the developers are not all located at the same geographical location, or are not located in close geographical proximity to each other. Likewise, if the development team is not located in close geographical proximity to the customer similar issues can result.

Geographical dispersion leads to various issues that do not exist when everyone is located at the same site, or, at least, are located relatively close to each other (e.g., in the same city or in two cities that are not far apart). Distributed development typically makes communication more difficult, because people are not able to interact at the same time and/or same place. Even if communication is not harder, distributed development requires special supporting tools, technologies, and communication mechanisms in order to address the unique requirements and characteristics of such an environment.

In distributed development environments, the Customer Interaction, Team Communication, Face-to-Face, and Documentation assumptions may not hold. The first three assumptions presume that it is very easy for developers to interact with each other and with customers. In fact, the Face-to-Face assumption assumes that developers and customers are all together where they can meet face-to-face – that they are co-located – since agile developers believe this is the most productive way to interact.

Geographical distribution makes interactions harder because of varying work schedules, differences in time zones, and because developers and clients cannot always see each other's reactions, and share ideas as flexibly and as clearly. The emphasis on co-location in agile processes does not fit well with the drive by some industries to realize globally distributed software development environments. Differential labor costs in other regions or other countries may motivate customers to employ offshore developers, or may motivate developers to use offshore labor. Development environments in which team members and customers are physically distributed may not be able to accommodate the face-to-face communication advocated by agile processes. In such cases, one can at least approximate face-to-face communication using technologies such as video-conferencing, chat and on-line whiteboards, conference calls, etc., but these technologies can be expensive and are not always as effective as one would hope.

Face-to-face communication can be as important in distributed environments as non-distributed ones. Such meetings must be planned in advance to ensure that all involved can participate and that the discussions are effective and not too time consuming. One can use such face-to-face meetings as major synchronization events in which geographically dispersed developers (1) are made aware of the progress made by others and (2) discuss plans for further evolving the product. In between such meetings, documentation (beyond code) may become the primary form of communication, with e-mail, chat, and video-conferencing technologies supplementing.

Good documentation of requirements and designs, produced and maintained in a timely manner, is essential to ensure that the distributed team members all maintain the same vision of the product to be built. This should not be interpreted as a requirement to document or model all aspects of software. Documentation and models should be created and maintained only if they provide value to the project and project stakeholders.

Agility is not always possible if communication is restricted to exchange of formal documentation due to legal reasons or due to world-wide distributed development. In these cases, only elements of agile process can be introduced locally, with formal processes being used to coordinate the larger, distributed project.

Unfortunately, in distributed environments especially, documentation is even more important because of differing time and place work activities, and different people and teams simultaneously and sequentially working on the same project. Documentation becomes more important because of the limited ways in which developers and customers can interact  The focus on minimizing documentation thus creates limitations in how well distributed development can be done following agile processes.

## *Limited Support for Subcontracting*

Outsourcing of software development tasks to subcontractors is often based on contracts that precisely stipulate what is required of the subcontractor. Subcontracted tasks have to be well-defined in the cases where subcontractors have to bid for the contract. In coming up with a bid, a subcontractor will usually develop a plan that includes a process – with milestones and deliverables – in sufficient detail to determine a cost estimate. The process could follow an iterative, incremental approach, but the subcontractor will likely have to make the process predictive by specifying the number of iterations, and the deliverables associated with each iteration, in order to compete. Because of this, the Customer Interaction, Team Communication, Face-to-Face, Documentation and Changing Requirements assumptions may not hold when work is subcontracted in a project.

As discussed above, the first three assumptions presume that developers and customers are all co-located so they can have face-to-face interaction whenever needed. It may not be possible to co-locate subcontractors with developers and customers. In these cases, the same issues that were identified for distributed development exist for subcontracting as well.

By requiring subcontractors to co-locate with the primary developers and the customer, these issues can be addressed.

As was discussed above, the documentation assumption states that documentation (other than actual program code) should only be created when absolutely necessary. In subcontracting, as was described for distributed development environments, documentation is important because people and teams who do not work together on a day-to-day basis must communicate and provide information so that others (other subcontractors, the main developers, the customer, etc.) can interact with what has been done and evaluate its acceptability within the project.

Given the greater "distance" between the main developers and subcontractors, and between the subcontractors and the customer, the assumption that documentation is not so important is easily seen to be invalid.

There is not much an agile development organization can do to address this issue other than to increase its documentation, or to require subcontractors to co-locate with them.

The changing requirements assumption states that requirements always evolve. However, subcontractors typically have won an award to develop software for a fixed set of requirements. If requirements change frequently, the contract has to change frequently, and this can lead to significant cost increases, since contracts typically state that there will be extra charges for each change to the contract. The basis of the contract used by agile developers and that of subcontractors is fundamentally different, since one assumes changing requirements and the other assumes a fixed set of given requirements.

In order to address this issue, it is possible that contracts can be written that allow a subcontractor some degree of flexibility in how they develop the product within time and cost constraints. This is certainly possible if the subcontractor has a good track record and can be trusted by the contracting company to develop a product that meets the contracting company's needs. A contract supporting agile development in the subcontractor environment might ought to consist of two parts:

- Fixed Part: This part defines (1) the framework that constrains how the subcontractor will incorporate changes into the product (e.g., cost- and time-based criteria for accepting or rejecting changes to the Variable Part (see below) of the contract, (2) the activities that must be carried out by the subcontractor (e.g., quality assurance activities), and (3) requirements that are to be considered fixed and deliverables that must be delivered.

- Variable Part: This part defines the requirements and deliverables that can vary within the boundaries defined in the Fixed Part. This part can evolve within the constraints defined in the Fixed Part. At the time the contract is signed, a description of prioritized deliverables and requirements should be included.

## Limited Support for Development Involving Large Teams

Large teams often have many sub-teams of specialists, and these may exist at different geographically-distributed locations. Large teams typically focus on very large projects, where a large amount of human resources are needed for solving the project's problems. Because of these issues, large teams require more interactions among their members and a higher degree of focus in order to manage them. In these environments, the Customer Interaction, Team Communication, Face-to-Face, and Documentation assumptions may not hold.

The size of teams can limit the effectiveness and frequency of face-to-face interactions. Agile processes support process "management-in-the-small" in that its coordination, control, and communication mechanisms are applicable to small to medium sized teams. With larger teams, the number of communication lines that have to be maintained can reduce the effectiveness of practices such as informal face-to-face communications and review meetings. Large teams require less agile approaches to tackle issues particular to "management-in-the-large".

There is not much that can be done to address this assumption other than to attempt to minimize the size of the team and to maximize the interaction that occurs, while at the same time not allowing the amount interaction to overwhelm the developers and the customer(s).

With large teams, more documentation is inherently needed, simply for coordinating among the large number of team members. Given their belief that any documentation other than code is to be minimized, agile development processes provide limited support for development involving large teams.

Traditional software engineering practices that emphasize documentation, change control and architecture-centric development are more applicable for large teams. This is not to say that agile practices are not applicable in such environments. There may be opportunities for large teams to use agile practices, but the degree of agility possible may be less than that found in smaller projects. For instance, the large overall team may have strict requirements for documentation, but, within this, it may be possible for small teams to apply agile development methods while they work on their project. After the project is completed, or at certain time intervals, the team may document certain aspects of the project so as to be in line with the large team's requirements. This would allow most of the work to be done in an agile manner, and only at the end (or other specified points) to produce required documentation.

## *Limited Support for Building Reusable Artifacts*

Reusable artifacts are code and other components (analysis and design documents, patterns, etc.) that can be reused from one project to another, in their entirety or at least in a major part. In order to create components that are reusable, a big-picture view must be taken while they are being developed, rather than simply focusing on the current application. What other types of systems / applications might be able to benefit from this component? How many different ways might one want to use it? What are the requirements of the domain, in contrast to simply this application in the domain? These are a few of the questions that must be asked when thinking about making components reusable and more general-purpose. When developing reusable artifacts, agile development's Documentation, Quality Assurance, Application-Specific Development, and Continuous Redesign assumptions may not be valid.

If documentation other than actual code is minimized, it may be harder to determine when and where a given artifact can be reused. Additional documentation may be needed to help indicate the reuse possibilities for an artifact. In order for agile processes to support development of reusable artifacts, they may need to increase the amount of documentation created.

Agile processes such as Extreme Programming focus on building software products that solve specific problems. Development in "Internet time" often precludes developing generalized solutions even when it is clear that this could yield long-term benefits. In such an environment, the development of generalized solutions and other forms of reusable software (e.g., design frameworks) is best tackled in projects that are primarily concerned with the development of reusable artifacts. This separation of the product-specific development environment from the reusable artifact development environment is a primary feature of the reuse-oriented framework called the *Experience Factory* developed by researchers at the University of Maryland at College Park [5]. The wide applicability of a reusable artifact requires that the process used to build the artifact emphasize quality control because the impact of low quality (in particular, severe errors) is as wide as the number of applications that reuse the artifact. On the other hand, timely development of reusable artifacts is desirable.

Continuous redesign is difficult when not developing application-specific artifacts. The opportunity for customer feedback is lessened, and thus the improvements in quality and design are reduced. In order to address this issue, agile developers must put in place specific processes that are intended to obtain this type of feedback so that the design and quality of the reusable artifacts can be enhanced.

It seems apparent that agile development does not naturally fit well for building reusable artifacts. However, with some careful attention, and some key adjustments made to agile processes, as mentioned above, it may be possible to successfully adapt and apply agile processes to development of reusable artifacts.

## *Limited Support for Developing Safety-Critical Software*

Safety-critical software is software where people's lives, health, or safety may be compromised if the quality of the software is not extremely high. Some examples include aviation control software, and software/firmware to control x-ray machines. In these types of environments it is important to know that software has been tested extensively, and has been designed to guarantee that there will not be failures that affect the ability to correctly and safely use and control the machinery. It is not acceptable for a machine to be allowed to give doses of x-rays that would be fatal to the patient receiving them, or for a pilot to be unable to fly the airplane because of software failure, for instance. In situations like these, the Documentation, Quality Assurance, and Continuous Redesign assumptions of agile development may not be valid.

Formal specification, rigorous test coverage, and other formal analysis and evaluation techniques included in software engineering approaches provide more robust, but also more expensive, mechanisms to tackle the development of safety- or business-critical software. These approaches can more reliably "guarantee" that appropriate tests have been run, and code has been analyzed, so that developers and users are confident in the safety and reliability of the system.

Applying some agile evaluation practices to such software can also be beneficial. For example, (1) test-first approaches requires one to define unit tests before writing code, (2) the early production of working code supported by the iterative, incremental process structure of agile processes supports exploratory development of critical software in which requirements are not well-defined, and (3) pair-programming can be an effective supplement to formal reviews.

Therefore, it can be assumed that agile and formal software development are not incompatible, but can be combined when needed: Formal techniques may be used in combination with agile processes to handle critical pieces of the software to increase quality and confidence.

## *Limited Support for Developing Large, Complex Software*

Large, complex software is software that includes large amounts of code (many hundreds of thousands, millions, etc., of lines) and/or may involve very intricate interrelationships between the various parts of the system to ensure data integrity and to make certain that all parts of the system interact reliably and run as intended. Development of large, complex software generally requires a higher degree of management control and a greater amount of more "formalized" processes to make sure everything fits and works together, and is runs reliably. The Documentation, Quality Assurance, Iteration, and Continuous Redesign assumptions of agile development may not be valid in these situations.

As was discussed above regarding large development teams, when developing large, complex software, it is likely that there is an increased need for documentation. This is necessary for simply documenting the larger set of requirements, features, and design decisions, as well as for providing a knowledge base for the larger teams that are likely to be working on such systems. Focusing almost exclusively on the code for the documentation can lead to a serious lack of understanding about the system, and the more difficult task of training new team members during and after the project is completed. If agile developers take the conscious effort to document key decisions, designs, etc., then this limitation may be able to be avoided.

Likewise, the assumption that informal testing and reviews can ensure the required level of quality in large complex systems is probably not valid. If the agile approach of creating tests before writing code (test-first) is carried out, and the process used in coming up with these tests is thorough and well-documented, then there may not be a problem. However, this needs to be ensured in order for quality in large, complex systems to be maintained.

The Iteration assumption may not be valid, either, when developing large, complex software because there may be systems in which functionality is so tightly coupled and integrated that it may not be possible to develop the software incrementally. In these cases an iterative approach in which code is produced in each iteration can still be used, but the code produced in each iteration will include all the pieces in various states of incompleteness.

Finally, the assumption that code refactoring removes the need to design for change may not hold for large complex systems in particular. In such software there may be critical architectural aspects that are difficult to change because of the critical role they play in the core services offered by the system. In such cases, the cost of changing these aspects can be very high and therefore it pays to make extra efforts to anticipate such changes early. The reliance on code refactoring (an application of the Continuous Redesign assumption) could also be problematic for such systems. The complexity and size of such software may make strict code refactoring costly and error-prone. Models can play an important role here, especially if tools exist for generating significant portions of the code from the models. This view of models as the central artifacts for evolving systems is at the heart of the Object Management Group's (OMG) Model-Driven Architecture (MDA) approach [28].

# 5. Open Questions, Conclusions, and Future Work

This paper has discussed claims made by agile developers, and some of the underlying principles and assumptions upon which agile development proceeds. Some of these assumptions have been questioned, and implications discussed. Some assumptions may always be true, but in other cases, these assumptions could lead to situations where agile development may not be applicable, or even where agile development may fail. In any case, there are a variety of questions that remain open and future work that needs to be done regarding agile development.

## *Open Questions*

While advances in software technologies and development tools have helped launch new generations of software products, it is also the case that new generations of software products drive the development of more sophisticated development infrastructures. It seems natural to assume that development might become more efficient and effective as the development infrastructure becomes more sophisticated. It would seem that development processes should improve over time as they adapt to the increasing sophistication of the development infrastructure. This raises the following open questions related to software development infrastructures and agile and "non-agile" processes:

- Do non-agile processes have a lot of 'overhead' because of the 'less-sophisticated' development infrastructure that existed at the time the processes were developed?
- Do agile processes work well because of the more sophisticated infrastructure that currently exists (e.g., component/class libraries, design frameworks, fast incremental compilers)?
- Would agile processes work so well if this infrastructure were not in place?
- What aspects of this infrastructure are key to making agile processes successful, and what aspects of agile processes themselves are responsible for their success? (The "nature-nurture" question.)

Answers to the above questions are not easy to obtain, but obtaining them can lead to a deeper understanding of development processes and their evolution.

## *Need for Empirical Studies*

While it appears that there have been many software development project successes based on agile processes, so far most of these success stories have only anecdotal evidence. For a more conclusive assessment of these new techniques, a sound scientific evaluation based on a statistically significant number of comparable case studies would be necessary. This could not only help one better understand unsolved and pressing problems in software engineering, but would also allow project managers to guide their decisions on process selection in a better way. It is invaluable to have hard numbers and data upon which to base our decisions about whether to adopt agile approaches to software development or not. Therefore, it is necessary to collect and analyze data about projects that have used agile processes. A first such step was done in [32]. Such studies will lead to a better understanding of how agile processes work, how they differ from "non-agile" processes, and under what conditions agile processes are applicable and are most successful.

Empirical data comparing the effectiveness and limitations of agile and non-agile approaches would greatly enhance our understanding of the true benefits and limitations of agile processes. In this paper we presented a list of limitations derived from our analysis of principles and assumptions underlying agile processes. It appears that certain domains are more amenable than others to agile development processes. Among them are Internet application domains, in which there are significant time-to-market pressures and the costs of upgrading to the next release are minimal. However, it also appears that companies that develop long-lasting, large, complex systems may not be able to use agile processes in their current form.

## *Spectrum of Development Approaches*

In general, some aspects of a software development project can benefit from an agile approach while others can benefit from a less-agile or more predictive approach. From this perspective, practical software development processes can be created by drawing techniques from agile as well as traditional approaches, rather than considering "agile" and "traditional" as discrete process classification points. Some projects can benefit from techniques that are more purely predictive, plan-based, "traditional" processes in which the process steps are defined in detail early in the project, and project goals remain relatively stable throughout the execution of the process. At the same time, these projects may also benefit from techniques that are more "agile" in which process steps and project goals are dynamically determined based on analyses of (1) experiences gained with previously executed process steps, (2) similar experiences gained outside of the project, and on (3) changes in the requirements and development environment. From this perspective, the agility of a process is determined by the degree to which a project team can dynamically adapt the process based on changes in the environment and the collective experiences of the developers.

Barry Boehm [10], in his analysis of agile practices, has proposed a process spectrum that is based on the degree of flexibility one has in developing process plans. Another way of looking at development processes might be in matrix form, with "agile" characteristics listed across one dimension and "traditional" ones listed across the other. The actual process used would be a combination of the characteristics selected from each of the two dimensions. This approach would fit in the vein of method engineering [16] where the specific processes and techniques that are desired for a project are selected from a catalog (method base) of available options.

Most agile process practices are adaptations of practices that have been touted by methodologists over the last two decades and that can be found in more rigorous "traditional" processes. This has been recognized by agile process advocates who point out that the differences lie not in the individual practices, but in how they are put together. The cobbling together of best practices to create processes that fit a development environment's values and development goals has been advocated by a number of methodologists and has resulted in at least one tailorable process framework, known as OPEN [16]. In this light, agile processes can be viewed as reference points along a spectrum of processes by those seeking processes that have the values embodied in the agile processes.

Practical processes lie somewhere in between the purely agile and purely predictive extremes of the process spectrum. Current agile processes are close to the purely agile end of the spectrum, but they are not purely agile because they provide a process framework that constrains the form of processes that developers must follow. For example, most published works on agile processes stipulate an iterative, incremental process and advocate practices such as test-first code development, pair-programming, and daily review meetings with particular formats.

## *Conclusions*

It is important to be aware that agile development approaches are built on many, possibly implicit, assumptions, and that these assumptions are probably not appropriate for all organizations or development projects. When the assumptions made by agile development methods are not in alignment, or even directly conflict, with those of the organization, managers in charge of development need to take steps to adapt the agile development process if such an approach is adopted, or be confident in choosing a "traditional" approach, knowing that it will better fit their environment. If this is not done, an agile development approach may very likely provide less than desirable results because of the limitations that result from these assumptions.